\documentclass[traditabstract]{aa}

\usepackage{epsfig,amsmath,amssymb}

\def\ga{\,\hbox{\hbox{$ > $}\kern -0.8em \lower 1.0ex\hbox{$\sim$}}\,}
\def\la{\,\hbox{\hbox{$ < $}\kern -0.8em \lower 1.0ex\hbox{$\sim$}}\,}
\def\beq{\begin{equation}}
\def\eeq{\end{equation}}

\titlerunning{Accretion driven turbulence and ion-neutral friction 
in self-gravitating filaments}
\authorrunning{Hennebelle}

\begin{document}

\title{Ion-neutral friction and 
accretion-driven turbulence in self-gravitating filaments}

\author{Patrick Hennebelle\inst{1,2}, Philippe Andr\'e \inst{1}}

\institute{
Laboratoire AIM, 
Paris-Saclay, CEA/IRFU/SAp - CNRS - Universit\'e Paris Diderot, 91191, 
Gif-sur-Yvette Cedex, France \\
\and
LERMA (UMR CNRS 8112), Ecole Normale Sup\'erieure, 75231 Paris Cedex, France}

\abstract{Recent Herschel observations have confirmed that filaments are ubiquitous 
in molecular clouds and suggest that irrespectively of the column density, there is a characteristic width of about 0.1 pc whose
physical origin remains unclear.
We develop an analytical model that can be applied to self-gravitating  accreting filaments. 
It is based on one hand on the virial equilibrium of the central part of the filament and on the other hand 
on  energy balance between the turbulence driven by accretion onto the filament and 
 dissipation. We consider two dissipation mechanisms the turbulent cascade and  the ion-neutral friction. 
Our model predicts that the width of the filament inner part is almost independent of the 
column density and leads to values comparable to what is inferred observationally 
if dissipation is due to ion-neutral friction. On the contrary turbulent dissipation leads 
to a structure that is bigger and depends significantly on the column density.
Our model provides a reasonable physical explanation which could explain the observed 
filament width when they are self-gravitating. It predicts the correct order or magnitude though 
 hampered by some uncertainties.}


\maketitle

\section{Introduction}
With the recent observations performed with $Herschel$ (e.g. Andr\'e et al. 2010, Molinari et al. 2010), it has become 
clearer that filaments are ubiquitous in molecular clouds and that they may play a central 
role in star formation.
While the exact influence filaments may  have on the star formation 
process remains to be clarified, it is important to understand 
the properties of these filaments since they are a direct consequence 
of the physics at play within molecular clouds. In this respect, 
a particularly intriguing observational result has been found by 
Arzoumanian et al. (2011) who showed that the central widths
 of the interstellar filaments
 have a narrow distribution that is peaked around a value of 0.1 pc. 
Moreover this characteristic  
width does not depend on the column density within the 
filament. Since it is believed that turbulence is important in 
molecular clouds and largely triggers their evolution, this result 
is at first sight counterintuitive since turbulence is  
responsible in a great variety of contexts 
for producing scale-free powerlaw distributions. 
Indeed, this suggests that  there is probably a physical 
process involved in setting this distribution which unlike 
turbulence presents a characteristic scale.

 A recent proposal made by Fischera \& Martins  (2012, see also Heitsch 2013, Gomez \& V\'azquez-Semadeni 2013) 
is that it may result from self-gravitating equilibrium. Indeed by solving 
hydrostatic equilibrium in an isothermal filament, Fischera \& Martins (2012) show that 
the filament width does not vary significantly and remains at a scale close to the observed 
0.1 pc. While this explanation is appealing, a few questions arise. First it assumes the existence of some confining 
pressure outside the filament whose nature remain to be specified. Second it fails to explain why the gravitationally 
unstable filaments which are collapsing also present this typical width.

In this paper, we explore the idea that the typical width of a
self-gravitating filament is due to 
the combination of accretion-driven turbulence onto the filament,
as suggested by the recent velocity dispersion measurements
 of Arzoumanian et al. (2013),  
and to the dissipation 
of this turbulence by ion-neutral friction which indeed  has 
a characteristic scale. 
 We stress that although ambipolar diffusion is considered here, the 
underlying idea is totally different  from the classical magnetically 
regulated star formation (e.g. Shu et al. 1987) in which the magnetic field 
is envisaged as the dominant support and ambipolar diffusion as the process 
through which the support can be circumvented. In the present picture clouds are 
typically supercritical. Note also that we do not address the reason why filaments 
form. As discussed in Hennebelle (2013) this is may be due to the shear of the turbulence 
with  possibly further amplification of the anisotropy by gravity for the most massive 
of them.

The plan of the paper is as follows. In the second section, 
we present the various assumptions and physical processes 
used in our simple model. The third section describes the results and
the fourth section concludes the paper.

\section{Model and Assumptions}

\subsection{Characteristics of the filament} 
Herschel observations (eg. Palmeirim et al. 2013) 
suggest that the typical average structure of  self-gravitating filaments 
is constituted by  $i)$ a central cylinder of nearly uniform density
$\rho_f$ and radius $r_f$, $ii)$ an envelope whose 
density profile is  $\propto r^{-2}$. This radial structure 
is reminiscent of many self-gravitating objects 
such as Bonnor-Ebert spheres. More precisely, filaments that 
are collapsing in a self-similar manner are expected to present 
an envelope with a profile $\propto r^{-2/(2-\gamma)}$ where 
$\gamma$ is the adiabatic index of the gas (Kawachi \& Hanawa 1998).
As it is likely that self-gravitating filaments are indeed collapsing 
in a way not too different from, although not identical to,
 a self-similar collapse, assuming an $r^{-2}$ profile is thus a 
well motivated assumption, both from observations and theory. 
One important difference with such  self-similar solutions  
however is that the central density plateau does not seem 
to be shrinking with time.

If $L$ is the length of the filament,  the mass 
of the central part is obviously 
$M_f = \pi \rho_f r_f^2 L = m_f L$
where $M_f$ is the  mass and $m_f$ the mass per unit length.
The total mass of the central part plus the surrounding 
envelope is 

\begin{eqnarray}
M_{\rm tot} = M _f [ 1 + 2 \ln ({r_{ext}/r_f}) ] \equiv M _f \mu ({r_{ext}/r_f}),
\label{mass_fil}
\end{eqnarray}
where $\mu(x)=1 + 2 \ln(x)$.
It has been assumed that 
the density outside the filament is $\propto 1/r^2$
and  that the filament stops at some radius $r_{ext}$.
Note that below we assume  $r_{ext}  \simeq L/2$.

\subsection{Gravitational potential within the filament}

The gravitational potential, in the radial direction is  obtained  by the 
Gauss theorem



\begin{eqnarray}
\label{grav_pot}
r<r_f, \; \phi(r)  &=& \pi G \rho_f r^2 = G m_f . \\
\nonumber
r>r_f, \; \phi (r) &=&  G m_f     \left( 1 + 2  \log(r/r_f)  + 2 (\log(r/r_f))^2   \right), \\
& \equiv &  G  m_f  ( 1+ {\cal G} (r/r_f) ). 
\nonumber
\end{eqnarray}
where ${\cal G}(x)= 2 \ln(x) + 2 (\ln(x))^2$.

\subsection{Magnetic field}
As the ion-neutral friction dissipation depends on 
the magnetic field, it is necessary to 
know its dependence.

We  proceed in two steps. First,  we 
discuss the expected value of the magnetic field 
in the parent clump, $B_c$.
Second,  we
infer the value of the magnetic field in 
the filament $B_f$ from the value of $B_c$. 

\subsubsection{Magnetic field in the parent clump}

The magnetic field in the clump is assumed to be proportional to the 
square root of the density $B_c = B_0 \sqrt{\rho_c / \rho_0}$ as indeed 
observed (e.g.  Crutcher 1999). Typical values 
are $B_0 \simeq 25 \mu$G and $n_0 \simeq 10^3$ cm$^{-3}$
where $n_0=\rho_0 / m_p$.
In the following we will use the 
$v_{A,0} = B_0 / \sqrt{4 \pi \rho_0} \simeq 1$ km s$^{-1}$ as 
a fiducial value.

Note that the magnetic field dependence is still debated
 and there are alternative choices. First, as suggested 
by Basu (2000), the magnetic field could indeed scale as  
$\sigma \sqrt{\rho}$, where $\sigma$ is the velocity dispersion,
 rather than just as $\sqrt{\rho}$. Second 
Crutcher et al. (2010) now favor $B \propto \rho ^{2/3}$. These 
relations do not represent large variations and would therefore 
not affect our results very significantly.

\subsubsection{Magnetic field in the filament}

To link the magnetic field in the filament to the magnetic field in 
the parent clump, we proceed as follows. 
First we assume that the magnetic field is perpendicular to the filament. 
This configuration is well supported by  observations in massive
filaments as  Taurus (Heyer et al. 2008, Palmeirin et al. 2013) or DR21 (e.g. Kirby 2009), and is also natural on  
physical grounds as the gas is expected to accumulate preferentially along 
the field lines.  This implies that at least a fraction of the gas accreted 
by the filament, is not impeded by the magnetic Lorentz force. There may  also be gas accreted 
perpendicularly to the field lines, which is therefore likely slowed down by magnetic 
pressure. However if the field is strong, then most of the material is presumably 
channeled along the field lines while if the field is weak it   also has a weak influence.

Second, we assume flux freezing which at these scales is a reasonable assumption. 
This implies that the magnetic field in the filament is simply the magnetic 
field in the clump compressed along the direction perpendicular to the field 
and the filament axis. Thus $B_f \simeq B_c \times \eta L/r_f$ since the matter 
that is inside the radius $r_f$ comes from a distance comparable to the
clump's size, $\eta L$, where $\eta$ typically varies with time between 0 and 1/2. 
A similar reasoning can be applied to get a relation between $\rho_c$
and $\rho_f$ since $M_f \propto \rho_f r_f ^2 L \simeq \rho_c \eta^2 L^3$.
We thus obtain $\rho_f = \rho_c (\eta L/r_f)^2$.
Combining the expression for $B_c$ obtained above with the latter expression, we get
$B_f = B_0 \sqrt{ \rho_f \over \rho_0}$
that is to say the magnetic field in the filament is also expected to be nearly 
proportional to $\sqrt{\rho_f}$ implying that the Alfv\'en velocity should 
remain nearly constant. 

 This relation is valid as long as flux freezing can be assumed. While this is 
a reasonable assumption  in the collapsing envelope, which is not magnetically supported, 
it is not the case in the central part, which is presumably close to equilibrium. Indeed, 
the typical ambipolar diffusion timescale  is given below by Eq.~(\ref{tau_diss_amb}).
For  densities on the order of $10^{4-7}$ cm$^{-3}$, the field is diffused 
in about 0.1-1 Myr, which is shorter than or comparable to, the accretion time of the
filaments. Moreover, as emphasized by Santos-Lima et al. (2012), turbulence also 
tends to diffuse out the field. 
 However since self-gravitating filaments are likely accreting, the magnetic flux cannot 
leak out far away since it is confined  by the infalling gas. Therefore, while the mean magnetic intensity
within the central part of such filaments is probably on the order of $B_f = B_0 \sqrt{ \rho_f \over \rho_0}$, it is 
likely that the magnetic field gradient is much reduced with respect to the ideal MHD case.

\subsection{Virial equilibrium}
The virial theorem is applied to the filament inner part of radius $r_f$.
The expression for a filament is (e.g. Fiege \& Pudritz 2000) 
\begin{eqnarray}
2 \int P dV + 2 E_{kin}^{cyl} =  W_{grav} + 2 P_{ext} V .
\label{virial_pasgen}
\end{eqnarray}
where  $W_{grav}$
is the gravitational term, $P$ and 
 $P_{ext}$ are the internal and external pressure and  $V$ the volume.
$E_{kin}^{cyl} = 0.5 M_f (2 \sigma_{1D} ^2)$ 
is the kinetic energy in the direction 
perpendicular to the filament axis and  
$\sigma_{1D}$
is the non-thermal one dimensional velocity dispersion.
This expression is however not strictly valid  since  our model filament is accreting.
Indeed further terms should 
be taken into account (see Goldbaum et al. 2011 and Hennebelle 2012)
which corresponds to the surface terms that do not cancel as it is 
usually the case. When the surface terms are taken into 
account, the virial expression becomes 

\begin{eqnarray}
{1 \over 2 } {\dot{M}^2} { dr_f^2 \over dM}  + 2 \int P dV + 2 E_{kin}^{cyl} =  W_{grav} + 2 (P_{ext} + P_{ram}) V .
\label{virial_gen}
\end{eqnarray}

At this stage we do not consider the 
influence that magnetic field may have on the equilibrium because it 
is not expected to change our conclusions  qualitatively.  In particular 
as discussed in the previous section, the magnetic gradient within the central 
part of the filament is probably smoothed due to ambipolar diffusion. Another 
complication arises because of the anisotropy introduced by the magnetic field being 
perpendicular to the filament axis, which would require a bi-dimensional analysis.

Using the different expressions obtained above we get
\begin{eqnarray}
{1 \over 2 } {\dot{m}_f^2 \over m_f} { dr_f^2 \over dm_f}  + 2 c_s^2 + 2 \sigma_{1D}^2 = G m _f + 2 {P_{ram} \over \rho_f},
\label{virial_comp}
\end{eqnarray}
where $P_{ram}$ is the ram pressure exerted by the incoming 
flow and where the pressure of the external medium has not been 
taken into account. The ram pressure will be estimated below.
For the sake of simplicity, we will also
use the simplified form of the virial equilibrium  
\begin{eqnarray}
2 \sigma_{1D} ^2 \simeq  G m_f.
\label{virial}
\end{eqnarray}

\subsection{Mechanical energy balance}

The  mechanical energy balance within the cylinder of radius $r_f$
leads to 
\begin{eqnarray}
 M_f { \sigma^2_{3D} (r_f) \over 2 \tau_{diss} } \simeq  
\epsilon _{eff} \dot{M} \left( \phi(r_{ext}) - \phi(r _f) \right).
\label{ener_balance}
\end{eqnarray}
Obviously the left hand-side is the dissipation which can be due 
either to the turbulence cascade or to the friction between 
ions and neutrals as described below. Note that it is assumed that 
the turbulence is isotropic which is why $\sigma_{3D}$ is used. 
The right hand-side describes the source of 
turbulence which is due to the accretion onto the central part of 
the filament (Klessen \& Hennebelle 2010). 
 The efficiency, $\epsilon _{eff}$, is not well known.
 Klessen \& Hennebelle (2010) proposed
that it can be related to the density contrast between the density 
of the incoming flow and the density of the actual gas in which 
energy is injected. In the present case, the accretion shock may not 
be clearly defined because of the turbulent nature of the flow.  
In any case, the present calculation remains
at this stage largely indicative. Below the value $\epsilon_{eff}=0.5$
is used because it leads to good agreement with the data. 
We stress that since our model remains indicative, $\epsilon _{eff}$ could 
also take into account various other uncertainties. 
For the sake of simplicity we have also ignored terms associated 
to the volume variation and the external pressure (e.g. Goldbaum et al. 2011)
as they do not modify the results substantially but makes the mechanical
energy balance much more complex.

\subsection{Accretion rate}
The accretion rate remains uncertain since it is difficult 
to infer observationally (see however Palmeirim et al. 2013 for an estimate 
in the case of the Taurus B211/3 filament). Here, we consider two different possibilities.
This will allow us to test the robustness of our conclusion.

\subsubsection{Gravitational accretion rate}
To estimate the accretion rate, we  assume that it can be computed 
from the density within the parent clump and the infall 
velocity. 
Assuming that the parent clump has a 
cylindrical radius of about $L/2$ and a length equal to $L$, we get
$\dot{M} =  \pi L ^2  \rho_c V_{inf}$.

The infall velocity is due to the gravitational field 
of the filament which is given 
by Eq.~(\ref{grav_pot}). We assume that 
the material which enters the clump at radius $r \simeq L/2$ has no initial velocity and we estimate 
the infall velocity at $r=L/4$ leading for $V_{inf}$ to
\begin{eqnarray}
V_{inf}   \simeq   
\sqrt{2  G  m_f}  \left( {\cal G} \left({L \over 2 r_f}\right) - 
 {\cal G} \left( {L \over 4 r_f}\right)  \right)^{1/2}.
\label{vinf}
\end{eqnarray}
Note that this is again a rough estimate but since the gravitational potential varies logarithmically 
with $r$, this estimate does not depend severely on these assumptions.

The density within the clump is also needed to get the accretion rate and we assume
 that $\rho_c= m_p n_c$ follows the Larson relations, 
(Larson 1981, Falgarone et al. 2009, Hennebelle \& Falgarone 2012)
 $m_p$ being the mass per particle
\begin{eqnarray}
n_c = n_0  \left( { R_c \over 1 {\rm pc}} \right)^{-\eta_d},\,\,
\sigma_{\rm 3D} = \sigma_0  \left( { R_c \over 1 {\rm pc}} \right)^{\eta},
\label{larson}
\end{eqnarray}
where $n_c$ is the clump gas density and $\sigma_{3D}$ the internal rms velocity. 
The exact values of the various coefficients remain somewhat uncertain. 
Originally, Larson (1981) estimated $\eta_d \simeq 1.1$ and $\eta \simeq 0.38$,
but more recent estimates (Falgarone et al. 2009) using 
larger sets of data suggest that $\eta_d \simeq 0.7$ and $\eta \simeq 0.45-0.5$.
For the sake of simplicity,  we use $\eta_d=1$ and take 
$n_0=1000$ cm$^{-3}$.

\begin{eqnarray}
\dot{M} \simeq   \pi L^{2-\eta_d}   \rho_0 (1 {\rm pc})^{\eta_d}
  \sqrt{2 G m_f}  \left( {\cal G} \left( {L \over 2 r_f} \right) - {\cal G} \left( {L \over 4 r_f} \right)  \right)^{1/2}
\label{accret_rate_grav0}
\end{eqnarray}

The typical accretion timescale is simply given by
$\tau_{accret}= {M / \dot{M}}$. With Eq.(\ref{accret_rate_grav0}), it is easy to show that 
$\tau_{accret} \propto \sqrt{\rho_f}$.

\subsubsection{Turbulent accretion rate}
As it is not clear what controls the accretion rates onto interstellar filaments, we
also consider a turbulent accretion rate constructed from 
the Larson relations as described in Hennebelle (2012).
\begin{eqnarray}
\label{dM_dt} 
\dot{M} &\simeq& {M_{\rm tot} \over \tau_c} \simeq {M_{\rm tot} \over 2 R_c/(\sigma_{3D}/\sqrt{3})} 
\simeq   \dot{M}_0 \left( { M_{\rm tot}  \over M_0} \right)^{\eta_{acc}} ,
\end{eqnarray}
where $\eta_{acc} \simeq 0.7-0.8$ depending on the exact choice of the parameter $\eta$ and $\eta_d$ that is retained.
We will adopt $\eta_{acc} \simeq 0.75$  as a fiducial value. 
We typically have $\dot{M}_0 = 10^{-3}$ M$_\odot$ yr$^{-1}$ for $M_0 =10^4$ M$_\odot$.

\subsubsection{Ram pressure}
The ram pressure which appears in Eq.~(\ref{virial_comp}) can be estimated as follows.
It is equal to the product of $V_{inf}(r_f)^2$ and $\rho_{in}=\dot{M}/(2 \pi r_f L V_{inf}(r_f))$,
where
 $\rho_{in}$ is  the density that is obtained assuming a constant accretion rate. Note that 
$\rho_{in} < \rho_{f}$ which implies that an accretion shock is connecting the infalling 
envelope and the central part of the filament. Since the above expression is assuming 
that the flow is isotropic and since we are assuming the structure of the flow rather 
than inferring an exact solution, this value remains hampered by large uncertainties and
 is certainly valid within a factor of a few. We have tested the influence of vaying 
the ram pressure by a factor of a few and found that it has a limited influence on the solution
 at low density while it has no influence at high density.

\subsection{Dissipation timescales}
 The dissipation timescale  to be used in Eq.~(\ref{ener_balance}) is 
a crucial issue. Here we emphasize two dissipation mechanisms, the turbulent cascade time 
and the ambipolar diffusion time. Quantitative estimates of these two timescales
have been recently estimated by Li et al. (2012) in turbulent two fluid MHD simulations. 
They found that under typical molecular cloud conditions both contribute but the latter 
dominates over the former.

\subsubsection{Dissipation by turbulent cascade}
First, we consider the standard turbulent cascade timescale, which is the 
crossing time of the system
\begin{eqnarray}
\tau_{diss,c} \simeq {2  r_f \over \sigma _{1D}}
\label{tau_diss_f}
\end{eqnarray}
The energy is  cascading to smaller and smaller scales until the size of the eddies
 reaches the viscous scale.

\subsubsection{Dissipation by ion-neutral friction}
Second, we investigate the dissipation induced by the ion-neutral
friction. Its expression has been first inferred by Kulsrud \& Pearce (1969, 
see also Lequeux 2005)
and is given by
\begin{eqnarray}
\tau_{diss, amb} = { 2 \gamma_{damp} \rho_i \over v_A^2 (2 \pi / \lambda)^2}
= { 2 \nu_ {ni} \over v_A^2 (2 \pi / \lambda)^2}, 
\label{tau_diss_amb}
\end{eqnarray}
where $v_A$ is the Alfv\'en speed, $\lambda$ is the wavelength assumed 
to be equal to $r_f$ and $\nu _{ni}$ is the ion-neutral coupling coefficient. 
The reason for choosing  $\lambda \simeq r_f$ is that 
if most of the energy is dissipated at a scale much smaller than 
$r_f$, then the relevant time would be the crossing or cascading time which would be
necessary for the energy to cascade from $r_f$. In this case, the timescale would be thus
similar to the turbulent cascade timescale discussed above.
The coefficient $\gamma_{damp}=3.5 \times 10^{13}$ cm$^3$ g$^{-1}$
s$^{-1}$
is the damping rate.
The ion density, $\rho_i$ is assumed to be $\rho_i = C \sqrt{\rho_n}$
where $C=3 \times 10^{-16}$ cm$^{-3/2}$ g$^{1/2}$.
For wavelengths,  $\lambda < \lambda_c= \pi v_A / (\gamma_{damp} \rho_i)$,
the critical wavelength,  
the Alfv\'en waves do not even propagate except at very small wavelengths when the ions 
and the neutrals are entirely decoupled. 
With $n\simeq 10^4$ cm$^{-3}$
and $v_A \simeq 1$ km s$^{-1}$, $\lambda_c \simeq 5 \times 10^{-2}$ pc.

Note that the expression stated by Eq.~(\ref{tau_diss_amb}) is strictly valid only for 
Alfv\'en waves. The corresponding expression for the compressible modes has been inferred
by Ferri\`ere et al. (1988). They are slightly more complex as it entails the 
angle between the field and the direction of propagation, but the order of magnitude is
not different.

\section{Results}
To infer the radius of the filament as a function of the central density, $\rho_f$, we 
have to combine Eqs.~(\ref{grav_pot}),~(\ref{virial}) and (\ref{ener_balance}) 
together with the accretion rate given by  Eq.~(\ref{accret_rate_grav0}) or Eq.~(\ref{dM_dt}) respectively. 

With the gravitational accretion rate (Eq.~\ref{accret_rate_grav0}), we obtain 
\begin{eqnarray}
{ \rho _f ^{1/2} r_f  \over \tau_{diss} } = {4 \over 3} \sqrt{2 \pi} \epsilon_{eff} \rho_0 
(1 {\rm pc}) \sqrt{G}  {\cal G}_1 (L/r_f)
\label{combi_grav}
\end{eqnarray}
where ${\cal G}_1(u) = \left(  {\cal G} (u/2) - {\cal G}(u/4) \right)^{1/2} 
{\cal G}(u/2)$.

With the turbulent accretion rate (Eq.~\ref{dM_dt}), we get the  relation
\begin{eqnarray}
{  (\pi \rho_f r_f^3) ^{1-\eta_{acc}} \over \tau_{diss}} = {4 \epsilon_{eff} \dot{M}_0  {\cal G}_2(L / r_f)  \over 3  M_0^{\eta_{acc}} (L/r_f) ^{1-\eta_{acc}} } 
\label{combi}
\end{eqnarray}
where ${\cal G}_2 (u) = \mu (u/2)^{\eta_{acc}} {\cal G} (u/2)$. 

\subsection{Dynamical equilibrium with turbulent dissipation}
Combining Eq.~(\ref{tau_diss_f}) with Eq.~(\ref{virial}), we get 
$\tau_{diss} = {2 \sqrt{2} \over \sqrt{\pi G \rho_f}}$
which in turn together with Eq.~(\ref{combi}) leads to the expression
\begin{eqnarray}
r_f = \rho_f ^{  \eta_{acc} -3/2 \over 3 (1-\eta_{acc}) } \left( { 8 \sqrt{2} \over 3 \pi ^{3/2-\eta_{acc}} \sqrt{G} } 
{ \epsilon_{eff} \dot{M}_0  {\cal G}_2 (L /  r_f)  \over M_0^{\eta_{acc}} (L/r_f) ^{1-\eta_{acc}} }  \right)^{1 \over 3(1-\eta_{acc}) }
\label{radius_turb}
\end{eqnarray}

\begin{figure} 
\includegraphics[width=9cm]{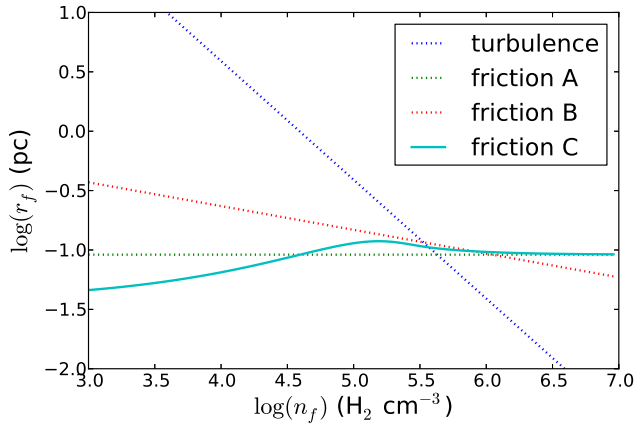}
\caption{Filament radius  as a function of filament density for four models.
The curve labeled ``turbulence'' shows result for 
a turbulent crossing time (Eq.~\ref{radius_turb}), the three curves labeled ``friction''
show results when ion-neutral friction  time is assumed to be the 
energy dissipation time (Eqs.~\ref{ambi_grav}-~\ref{ambi_turb}).
Frictions A and C used a gravitational accretion rate and 
Friction B a turbulent one.}
\label{radius_comp}
\end{figure}

For the canonical value $\eta_{acc}=0.75$, we find that $r_f \propto 1 / \rho_f$. That is to say, 
the typical filament radius decreases with the central density as 
displayed in Fig.~\ref{radius_comp} (see line labeled turbulence). Note that the value 
$\epsilon _{eff}=0.5$ has been used in this calculation. 
For larger values of $\eta_{acc}$
 we still get significant variations of  $r_f$ with $\rho_f$. For $\eta_{acc}=1$
we would even predict that the central density is independent of  $r_f$.
This is only for $\eta_{acc} \simeq 3/2$ that a filament radius independent of the central 
density is obtained. The gravitational accretion rate expression leads 
to a very similar expression with $r_f \propto 1/\rho_f$ and the corresponding expression is not given here 
for conciseness as the obvious conclusion is that this behaviour is incompatible with the 
filament width being nearly constant.

\subsection{Dynamical equilibirum with ion-neutral friction}
Combining Eq.~(\ref{tau_diss_amb}) and Eq.~(\ref{virial}),  
with  Eq.~(\ref{combi_grav}), we infer
\begin{eqnarray}
r_f = { 3 \pi ^{3/2} v_{a,0}^2 \over 2 \sqrt{2} \epsilon_{eff} \gamma_{damp} C \rho_0 (1 {\rm pc}) \sqrt{G} {\cal G}_1 (L/r_f)}.
\label{ambi_grav}
\end{eqnarray}
where we see that $r_f$ does not depend on $\rho_f$. 
That is to say the width of the filament does not change 
with its column density as suggested from the results of Arzoumanian et al. (2011). To test 
the robustness of this result, it is worth investigating what the turbulent accretion rate stated 
by Eq.~(\ref{dM_dt}) is predicting. The corresponding expression is 
\begin{eqnarray}
r_f = \rho_f ^{ \eta_{acc} -1/2 \over 1 - 3 \eta_{acc}} \left( { 2 \gamma_{damp} C  \over  3 \pi^{3-\eta_{acc}} v_{a,0} ^2 }  
{\epsilon_{eff} \dot{M}_0  {\cal G}_2 (L / r_f)  \over M_0^{\eta_{acc}} (L/r_f) ^{1-\eta_{acc}} }  \right)^{1 \over 1-3 \eta_{acc} } 
\label{ambi_turb}
\end{eqnarray}
As can be seen for an accretion rate exponent $\eta_{acc}$ of the order of 
0.75, we find 
that $r_f \propto \rho_f ^{-0.2}$, which implies a very shallow dependence 
of the filament radius $r_f$. For a value of $\eta_{acc} =1$, we have 
$r_f \propto \rho_f ^{-1/4}$
which is still a shallow dependence as displayed by Fig.~\ref{radius_comp} where 
the two expressions stated by Eqs.~(\ref{ambi_grav}) and~(\ref{ambi_turb}) are displayed (labeled 
as friction A and B respectively). 

\subsection{A more accurate model}
\begin{figure} 
\includegraphics[width=9cm]{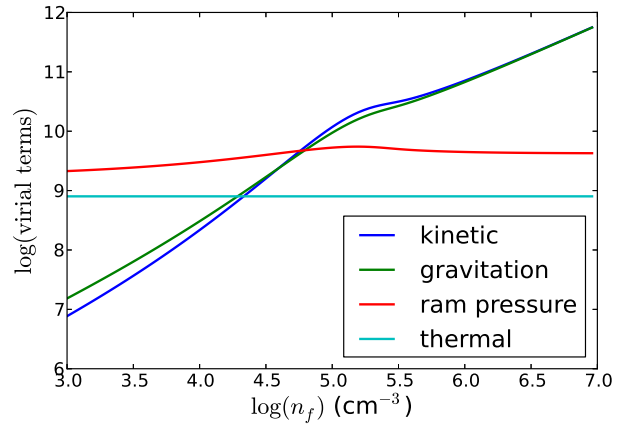}
\caption{Amplitude of the various terms which appear in the virial equilibrium
(Eq.~(\ref{virial_comp})). While at high density the filament equilibrium is due to 
the balance between gravity and velocity dispersion, it is due to the balance between 
ram  and thermal 
pressure at low density.}
\label{coef_vir}
\end{figure}

Finally, we investigate the solutions when the thermal support and 
ram pressure are considered as stated in Eq.~(\ref{virial_comp}). The 
corresponding curve is labeled as friction C. 
 Equation~(\ref{virial_comp}) is an ordinary differential equation in $r_f$,
which can be solved using a standard Runge-Kutta method. Since it is necessary 
to specify a radius and a density to start the integration, we have 
explored various cases. We found that for a large range of radii ($r_{start} \simeq 0.01-0.1$ pc) at low density, 
the solutions quickly converge towards the one that is presented here and for which the 
radius at $n=10^3$ cm$^{-3}$ is equal to about 0.05 pc. 

As can be seen more 
variability is introduced, particularly at low density where we see that the filament 
radius decreases at low density. In order to better understand the physical meaning of this 
solution we plot in Fig.~\ref{coef_vir}, the values of the various virial terms as a function of density.

While the equilibrium between gravity and turbulent support at high density is 
robust and independent of the choice of the boundary condition, $r_{start}$, the 
behaviour at low density is less robust and varies with it.  

It is important to stress a few points. 
First the ram pressure term which causes most 
of the variation remains uncertain since our model 
is not fully self-consistent in the sense that the 
density and velocity fields, although reasonable, are not 
proper solutions of the problem. Second in the low density 
regime, the filament is  not gravitationally accreting 
and it is likely that the validity of the model is questionable.

\subsection{Comparison of the two dissipation timescales}

\begin{figure} 
\includegraphics[width=9cm]{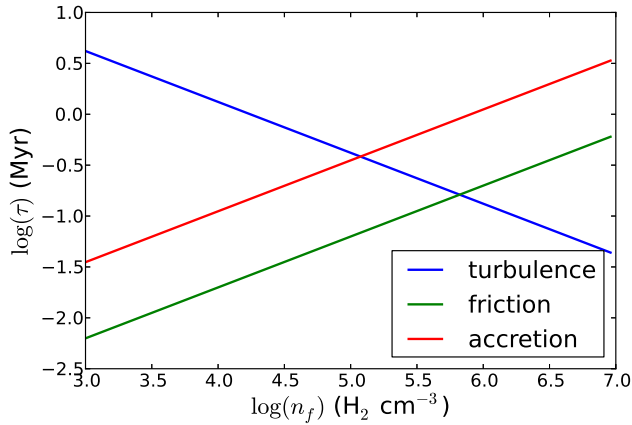}
\caption{Comparison between the turbulent dissipation and 
ion-neutral friction times as a function of density. 
Also shown is the accretion timescale. 
It scales exactly as the 
ion-neutral friction times.}
\label{time_comp}
\end{figure}

 It is enlighting to compare the values of the dissipation timescales
as a function of density for the filament radius corresponding to model 
B. As expected the turbulent dissipation timescale is much longer 
than the ion-neutral friction timescale for densities lower than 
$10^6$ cm$^{-3}$. It is also increasing with  density 
while the turbulent one is decreasing with density. This behaviour
is the very reason which explains the nearly constant width 
of the filaments in our model because as can be seen in Fig.~\ref{time_comp}, the 
accretion time present the same dependence.

To summarize, assuming that the relevant timescale for  energy 
dissipation within the central part of the filament is the turbulent 
crossing time, we find under reasonable assumptions for the accretion 
rate that the width changes significantly with density. This is 
because $\tau_{diss,c} \propto 1/ \sqrt{\rho_f}$. 
On the other hand, when we assume that the 
relevant timescale for energy dissipation is the ion-neutral friction time, 
we find that the width varies much less with the filament density. This
is because, $\tau_{diss,amb} \propto \rho_i \propto \sqrt{\rho_f} \propto \tau _{accret}$.
Since the relevant timescale is the shortest one, which corresponds 
to the smallest value of $r_f$, one expects  ion-neutral friction
to be the dominant mechanism for energy dissipation up to densities
equal to a few $10^6$ cm$^{-3}$ (see Fig.~\ref{radius_comp}). 

\subsection{Comparison with observations}

\begin{figure} 
\includegraphics[width=9cm]{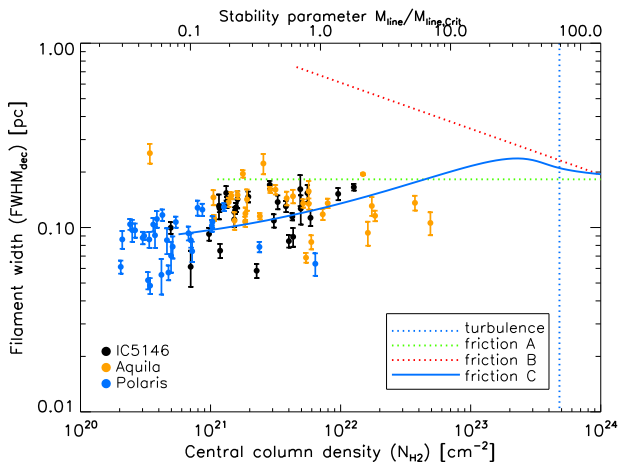}
\caption{Comparison between the models and the 
filaments width distribution (adapted from Arzoumanian et al. 2011).}
\label{data_comp}
\end{figure}

Finally, we confront the present models with the $Herschel$ observational 
result of Arzoumanian et al. (2011). Figure~\ref{data_comp} shows
 filament width as a function of filament column density. As can be seen 
the model based on ion-neutral friction and gravitational 
accretion (friction B and C) work very well. Note that the 
model based on turbulent dissipation  predicts a 
constant column density and a variable radius.

\section{Conclusion}
We have presented a simple model  to describe the evolution of
accreting  self-gravitating 
filaments within molecular clouds.  It assumes virial equilibrium 
between gravity and turbulence and mechanical energy balance
between accretion which drives turbulence in the filament and 
the dissipation of this energy.  We show that while 
dissipation based on turbulent cascade fails to reproduce
the narrow range of radius inferred from $Herschel$ observations, 
dissipation based on ion-neutral friction leads to 
a filament width that  depends only weakly on the filament density
and is very close to the $\simeq$0.1 pc value although our analytical approach is hampered 
by significant uncertainties.
We conclude that the combination of accretion-driven turbulence and ion-neutral friction
 is a promising mechanism to explain the structure of self-gravitating filaments and deserves 
further investigation. \\

\emph{Acknowledgments}
We thank the anonymous referee for a constructive and helful report.
We thank Doris Arzoumanian and Evangelia Ntormousi for discussions on this topic.
PH acknowledge the financial support of the Agence National pour la Recherche through the 
 COSMIS project.
This research has received funding from the European Research Council under the European
 Community's Seventh Framework Programme (FP7/2007-2013 Grant Agreement no. 306483 and no. 291294).

\end{document}